\documentclass[onecolumn,preprintnumbers,amsmath,amssymb]{revtex4}

\begin{document}


\title{Thermodynamics of a photon gas and deformed dispersion relations}
\author{Abel Camacho}
\email{acq@xanum.uam.mx} \affiliation{Departamento de F\'{\i}sica,
Universidad Aut\'onoma Metropolitana--Iztapalapa\\
Apartado Postal 55--534, C.P. 09340, M\'exico, D.F., M\'exico.}

\author{Alfredo Mac\'{\i}as}
\email{amac@xanum.uam.mx} \affiliation{Departamento de
F\'{\i}sica,
Universidad Aut\'onoma Metropolitana--Iztapalapa\\
Apartado Postal 55--534, C.P. 09340, M\'exico, D.F., M\'exico.}

\date{\today}

\begin{abstract}
We resort to the methods of statistical mechanics in order to
determine the effects that a deformed dispersion relation has upon
the thermodynamics of a photon gas. The ensuing modifications to
the density of states, partition function, pressure, internal
energy, entropy, and specific heat are calculated. It will be
shown that the breakdown of Lorentz invariance can be interpreted
as a repulsive interaction, among the photons. Additionally, it
will be proved that the presence of a deformed dispersion relation
entails an increase in the entropy of the system. In other words,
as a consequence of the loss of the aforementioned symmetry the
number of microstates available to the corresponding equilibrium
state grows.
\end{abstract}

\maketitle
\section{Introduction}

Due to its fundamental role in modern science Lorentz symmetry has
been subjected to some of the highest precision tests \cite{[1],
[2],ACLMM} that physics has performed. In spite of this severe
experimental scrutiny several quantum gravity models predict the
breakdown of Lorentz symmetry \cite{[3], [4], [5]}. Additionally,
it has been suggested an energy dependent speed of light as a
possible solution to the so--called GZK paradox \cite{[6]},
namely, the observation ultra high energy cosmic ray above the
expected GZK threshold for interaction of such cosmic rays with
the cosmic microwave background \cite{[6], [7]}.

 The possible detection of
these corrections has already been analyzed in several contexts.
Indeed, in the realm of interferometry it has been analyzed
resorting to first--order coherence proposals \cite{[8], [9]}, or
even higher--order coherence experiments, like the so--called
Hanbury--Brown--Twiss experiment \cite{[10]}, the modifications
upon the Standard Model that a Lorentz invariance violation could
have \cite{[11], [12], [13]}.

In the present work the consequences of a Lorentz invariance
violation will be considered in a very different approach. The
idea here is to introduce a deformed dispersion relation as a
fundamental fact for the dynamics of photons. Afterwards we
analyze the effects of this assumption upon the thermodynamics of
a photon gas. It will be shown that the breakdown of Lorentz
symmetry entails an increase in the number of microstates, and in
consequence a growth of the entropy, with respect to the case in
which Lorentz symmetry exists. Resorting to the case of a
non--ideal gas (the interactions among the particles cannot be
neglected) it will be proved that the presence of a deformed
dispersion relation could be interpreted as a repulsive
interaction. The specific heat, the entropy, and other
thermodynamic parameters are also calculated, and the consequences
of the breakdown of Lorentz symmetry are evaluated.
\bigskip
\bigskip

\section{Thermodynamical consequences of a deformed dispersion relation}
\bigskip

\subsection{Density of states and partition function}
\bigskip

As mentioned above several quantum--gravity models predict a
modified dispersion relation \cite{[3]}, the one can be
characterized, phenomenologically, through corrections hinging
upon Planck's length, $l_p$,

\begin{equation}
E^2 = p^2\Bigl[1 - \alpha\Bigl(El_p\Bigr)^n\Bigr].
\label{Disprel1}
\end{equation}

Here $\alpha$ is a coefficient, whose precise value depends upon
the considered quantum--gravity model, while $n$, the lowest power
in Planck's length leading to a non--vanishing contribution, is
also model dependent.

Consider now a photon gas in a container of volume $V$. Though the
momentum, as a consequence of the boundary conditions, is
quantized, we may assume, as is usual under these conditions
\cite{[14]}, a continuous momentum spectrum. At this point it is
noteworthy that this assumption has nothing to do with the
relation between energy and momentum, and in consequence the error
here introduced is the same that appears in the usual model for a
photon gas. Then the number of states in a certain volume of the
phase space is given by

\begin{equation}
\Sigma = \frac{1}{(2\pi\hbar)^3}\int\int d\vec{r}d\vec{p}.
\label{Numstates1}
\end{equation}

Our deformed dispersion relation entails that (here $E_p =
\sqrt{G/(c^5\hbar)}$ denotes the so--called Planck's energy)

\begin{equation}
p^2 = \frac{E^2}{c^2}\Bigl[1 -
\alpha\Bigl(E/E_p\Bigr)^n\Bigr]^{-1}. \label{Momentum}
\end{equation}

At this point we remember that Statistical Mechanics tells us that
the relation between energy and momentum, of the particles
comprising a gas, has a very important role in the evaluation of
the dependence of the pressure as a function of the energy density
\cite{[14]}. Indeed, assume that the momentum is a power, say $s$,
of the energy, and that there are $l$--space--like dimensions,
mathematically $p\sim\epsilon^s$, (here $p$ and $\epsilon$ are
momentum and energy, respectively), then the pressure $P$ and the
energy density, $u$, of the corresponding gas are related by the
expression $P=su/l$. The last expression entails that we must now
have a different state equation, i.e., a deformed dispersion
relation must change the thermodynamical properties of a photon
gas, since now the relation between momentum and energy is not
given by $p\sim\epsilon$.

\begin{equation}
\frac{dp}{dE}= \frac{1}{c}\Bigl[1 -
\alpha\Bigl(E/E_p\Bigr)^n\Bigr]^{-3/2}\Bigl[1 +
\alpha(n-1)\Bigl(E/E_p\Bigr)^n\Bigr]. \label{Dermomentum}
\end{equation}

Then, if the gas is in a container with volume $V$, we may cast
the number of states in the following form

\begin{equation}
\Sigma = \frac{4\pi V}{(2c\pi\hbar)^3}\int_0^{\infty}E^2\Bigl[1 -
\alpha\Bigl(E/E_p\Bigr)^n\Bigr]^{-5/2}\Bigl[1 +
\alpha(n-1)\Bigl(E/E_p\Bigr)^n\Bigr]dE. \label{Numstates2}
\end{equation}

At this point it is noteworthy to comment that (assuming $\alpha
>0$) the number of available states grows, with respect to the case
in which Lorentz symmetry is preserved, when a deformed dispersion
relation appears. In other words, the absence of the symmetry
implies that with respect to the corresponding equilibrium state
now more microstates are available. This last remark implies also
that we should not be surprised if the corresponding entropy
grows, clearly, here we mean a larger entropy compared against the
entropy of a gas under the same conditions, but with the Lorentz
symmetry untouched.

Let us now proceed to calculate the thermodynamical properties of
such a photon gas, and in order to do this let us remember that
the connection between the microscopic world and the thermodynamic
behavior is done by the partition function ${\cal Z}$ \cite{[14]},
i.e., the knowledge of this function allows us to deduce all the
thermodynamics of the corresponding system.

\begin{equation}
\ln {\cal Z} = -\frac{4\pi
V}{(2c\pi\hbar)^3}\int_0^{\infty}E^2\Bigl[1 +
\alpha(n+3/2)\Bigl(E/E_p\Bigr)^n\Bigr]\ln\Bigl[1 -
e^{-E/KT}\Bigr]dE. \label{Partition1}
\end{equation}

Explicitly, the logarithm of the partition function reads

\begin{equation}
\ln {\cal Z} = \frac{8\pi^5 V(KT)^3}{(2c\pi\hbar)^390}[1 +
\frac{\alpha}{2}(n+3/2)[(n+2)!]\frac{\xi(4+n)}{\xi(4)}(T/T_p)^n].
\label{Partition2}
\end{equation}

In this last expression $K$ is Boltzmann's constant, $T_p= E_p/K$
Planck's temperature, and $\xi(x)$ the so--called Riemann's zeta
function \cite{[15]}. Setting $\alpha =0$ everything reduces to
the usual case. Notice that the new contribution depends upon term
$(T/T_p)^n$, a fact that complicates the possible detection of
these effects, due to the value of $T_p$.
\bigskip

\subsection{Thermodynamical parameters}
\bigskip

 Recalling that the connection between internal energy and partition function is given by \cite{[14]}

 \begin{equation}
U = KT^2\Bigl(\frac{\partial \ln {\cal Z}}{\partial T}\Bigr)_V.
\label{Intenergy0}
\end{equation}

\begin{equation}
U = \frac{24\pi^5 V(KT)^4}{(2c\pi\hbar)^390}[1 +
\alpha(n+3/2)\frac{(n+3)!}{3!}\frac{\xi(4+n)}{\xi(4)}(T/T_p)^n].
\label{Intenergy1}
\end{equation}

 If in (\ref{Intenergy1}) we set $\alpha=0$, then we
recover the usual internal energy for a photon gas \cite{[14]}.
Also notice that the breakdown of Lorentz symmetry entails an
increase in the internal energy of our photon gas. Though the
internal energy is not directly a detectable parameter, there are
others variables, as the specific heat, that may be measured and
which are calculated from the internal energy.

We now proceed to calculate the state equation. This is easily
done if we remember that it can be deduced from the characteristic
function of the corresponding ensemble \cite{[14]}. For the
present case we have that in the Grand Canonical Ensemble the
characteristic function is given by

\begin{equation}
PV = KT \ln {\cal Z}. \label{Charfunction}
\end{equation}

This expression clearly show us that if we have $\ln {\cal Z}$
then immediately we obtain the state equation. Our previous
results imply that the pressure is given by

\begin{equation}
P = \frac{8\pi^5 (KT)^4}{(2c\pi\hbar)^390}[1 +
\frac{\alpha}{2}(n+3/2)[(n+2)!]\frac{\xi(4+n)}{\xi(4)}(T/T_p)^n].
\label{Pressure1}
\end{equation}

It is important to mention that the pressure grows, with respect
to the case in which Lorentz symmetry is present, the condition
$\alpha =0$ takes us back to the usual situation in which the
pressure behaves as $P\sim T^4$.

This last remark allows us to interpret the breakdown of Lorentz
symmetry as a repulsive interaction. Indeed, the presence of a
repulsive interaction (among the particles of a gas) entails the
increase of the pressure, compared against the corresponding value
for an ideal gas.

Let us explain this point deeper. A fleeting glimpse at the
cluster expanssion and its relation to the virial coefficients
\cite{[16]} clearly shows that the first correction to the ideal
gas state equation expressed in terms of the so--called virial
state equation ($PV/(NKT) =
\Sigma_{l=1}^{\infty}a_l(T)\bigl(N\lambda^3/V\bigr)^{l-1}$, $N$
denotes the number of particles) corresponds to a virial
coefficient that can be written, as a function of the potential
energy of interaction between the i--th and the j--th particle
$v_{ij}$ as

\begin{equation}
a_2 = -\frac{1}{\lambda}\int f_{12}d^3r_{12}.\label{2virialcoeff}
\end{equation}

Where $\exp\bigl\{-v_{12}/KT\bigr\} = 1 + f_{12}$, here $f_{12}$
is the two particle function (which is a function of the potential
$v_{12}$ between two particles), and $\lambda =
\sqrt{2\pi\hbar^2/mKT}$ is the so--called thermal wavelength
\cite{[16]}.

A repulsive interaction means that $v_{12}>0$, and in consequence
$f_{12}<0$, and therefore, $a_2>0$. If we introduce this condition
into the virial expression we obtain a pressure larger than that
corresponding to an ideal gas. In other words, the introduction of
a repulsive interaction among the particles comprising the gas
entails an increase of the pressure, compared to the pressure of
an ideal gas.

It is in this sense that we say that the loss of the symmetry
appears, at the bulk level, as the emergence of a repulsive
interaction, and in consequence, at least in principle, we could
detect some effects stemming, either from loop quantum gravity,
non--commutative geometry, etc.

One of the consequences of Lorentz symmetry can be seen casting
the pressure in terms of the energy density ($u= U/V$), i.e.
$P=u/3$. Indeed, if the relation between energy and momentum for a
bosonic particle in $l$--spacelike dimensions is $\epsilon\sim
p^s$, then the pressure becomes $P=\frac{s}{l}u$ \cite{[14]}.
Notice that for our case the aforementioned relation between
energy and momentum does not hold anymore, and in consequence the
expression for the pressure in terms of the energy density can not
have this form. In other words, the breakdown of Lorentz symmetry
can be, in principle, detected looking at the relation between
pressure and energy density, i.e., this kind of effects do appear
at the macroscopic level. For our case we have that

\begin{equation}
P = \frac{1}{3}u\Bigl[1
-\alpha(n+3/2)[(n+2)!]\frac{n}{3!}\frac{\xi(4+n)}{\xi(4)}(T/T_p)^n\Bigr].
\label{Pressure2}
\end{equation}

Setting $\alpha=0$ we obtain the usual result. As expected, the
corrections to the usual behavior are a function of the term
$(T/T_p)^n$.

Another interesting thermodynamical quantity is the entropy, $S$.
Since $S= PV/T + U/T$, and with all the results we may write

\begin{equation}
S = \frac{8\pi^5V(KT)^3}{90(2c\pi\hbar)^3}K\Bigl[4
+\alpha(n+3/2)(n+4)\frac{(n+2)!}{2!}\frac{\xi(4+n)}{\xi(4)}(T/T_p)^n\Bigr].
\label{Entropy1}
\end{equation}

A consequence of (\ref{Entropy1}) is that a reversible adiabatic
process is not anymore given by the condition $VT^3 = const.$, as
is usual \cite{[14], [16]}. Additionally, for a photon gas $3S =
T(\frac{\partial S}{\partial T})_V$. This is also lost if $\alpha
\not= 0$. Notice that the breakdown of Lorentz symmetry increases
the number of microstates available to the macrostate, see
(\ref{Numstates2}), and in consequence, our entropy grows, as
(\ref{Entropy1}) clearly displays, with $\alpha >0$.

Since the internal energy suffers a change in its dependence upon
the temperature (\ref{Intenergy1}) then the specific heat, in this
case at constant volume, must also show modifications.

Recalling that

\begin{equation}
C_V = \Bigl(\frac{\partial U}{\partial
T}\Bigr)_{V},\label{Specifheat0}
\end{equation}

\begin{equation}
C_V = \frac{24\pi^5V(KT)^4}{90T(2c\pi\hbar)^3}\Bigl[4
+\alpha(n+3/2)(n+4)\frac{(n+3)!}{3!}\frac{\xi(4+n)}{\xi(4)}(T/T_p)^n\Bigr].
\label{Specifheat1}
\end{equation}

It shall be no surprise that the value $\alpha =0$ reduces the
specific heat to its usual value. This is a measurable quantity
and in consequence we have obtained another parameter, which in
principle, could be employed in the experimental quest for
violations of Lorentz symmetry. The calculated specific heat is
larger than the corresponding parameter for an ideal gas, and this
result is compatible with our previous interpretation of the
breakdown of Lorentz symmetry as the appearance of a repulsive
interaction. Indeed, the specific heat at constant volume, in
regions where the repulsive interaction among the particles plays
the predominant role, is larger than the value for an ideal gas,
see table 3.1 on page 77 of \cite{[17]}. This can also be
understood if in the van der Waals state equation (in which $v$
denotes the volume per particle, $a$ is a parameter related to an
attractive interaction among the particles, whereas $b$ is related
to a repulsive one \cite{[14], [16]})

\begin{equation}
P = \frac{RT}{v- b} - \frac{a}{v^2}, \label{van der Waals}
\end{equation}
\bigskip
\bigskip

we impose the condition $a=0$, and calculate the corresponding
specific heat with $b>0$, a condition that can be rephrased in
terms of the repulsive interaction that appears among the
particles due to the fact that they are not point--like particles,
but do have a non--vanishing volume.

At this point it is noteworthy to mention that, as can be seen
from the previous expressions, the modifications, due to this kind
of breakdown, behave like $(T/T-p)^n$, and, in consequence, a
terrestrial experiment in this context is not feasible.
\section{Conclusions}
\bigskip

We have assumed from the onset the modification of the dispersion
relation, for photons, a condition stemming in many models that
try to quantize gravity. With this new relation the density of
states has been evaluated. It has been shown that the number of
microstates available to the corresponding equilibrium state
grows, compared to the case in which Lorentz symmetry is present.
As expected, the entropy becomes larger as an unavoidable
consequence of this kind of Lorentz violation (\ref{Entropy1}),
nevertheless, in the limit $T\rightarrow 0$ the entropy goes to
zero, as happens in the case in which Lorentz symmetry is present.
In other words, Nernst postulate is not violated if the symmetry
is broken.

Additionally, since the breakdown of Lorentz symmetry entails a
larger pressure (compared to the case of a photon gas in which
Lorentz symmetry is present), see (\ref{Pressure2}), then we may
interpret the breakdown of Lorentz symmetry, at least in the bulk
realm, as equivalent to the emergence of a repulsive interaction
among the photons. This interpretation is corroborated by the
ensuing form for the specific heat at constant volume. Indeed,
even for a simple case, for instance, van der Waals state equation
(in which the fact that the molecules do have a non--vanishing
volume can be considered as a repulsive interaction the one
appears when the distance between two particles is smaller than
the radius of the corresponding molecules) the specific heat may
become larger than $3/2$, the value of an ideal classical gas, see
\cite{[17]}, table 3.1 on page 77.

Concerning the detection of these effects let us mention that it
could be done, at least in principle, resorting to the specific
heat at constant volume (clearly, the specific heat at constant
pressure is also modified), or the pressure. Unfortunately, the
modifications to the usual case appear also as a function of
$T/T_p$, where $T_p\sim 10^{32}$ Kelvin is the so--called Planck's
temperature, and this is a very stringent condition,
experimentally.

Since the modification of the thermodynamical behavior, as a
consequence of a deformed dispersion relation, can be explained as
a result of the fact that the state equation of the gas depends
upon the relation between momentum and energy of the particles,
then it is readily seen that for a bosonic gas comprising massive
particles the thermodynamics will be also changed. In this
context, since the so--called Bose--Einstein condensation is a
purely bosonic effect, then we may wonder what happens to this
feature if we introduce the generalization to (\ref{Disprel1}) for
massive particles. The consequence is a modification of the
condensation temperature. In a quite similar spirit we may analyze
the situation of fermions, and for instance, the Fermi temperature
is modified. This last case yields an interesting situation, since
the Chandrasekhar mass-radius relation for white dwarfs is a
direct consequence of the fermionic statistics \cite{[14]}, and
hence we expect a modification of this relation due to the
breakdown of Lorentz. This last remark opens a new window in this
context, an astrophysical one connected to the observation of
white dwarfs. The corresponding results of this analysis will be
published elsewhere.

Finally, let us comment the coincidences and divergences of the
present work with some previous results in this direction, for
instance, \cite{[18], [19], [20]}. Though these aforementioned
papers do analyze the implications in the realm of statistical
mechanics of quantum--$\kappa$--Poincar\'e algebra \cite{[18]}, or
of a Generalized Uncertainty Principle \cite{[19], [20]}, they do
not consider the interpretation of the breaking of Lorentz
symmetry, at the bulk level, as the emergence of a
pseudo--interaction among the particles of the gas. This fact,
which is present even for massive particles \cite{[21]}, allows us
define two different classes of Lorentz symmetry breaking, i.e.,
those related to an attractive pseudo--interaction, and those
connected with a repulsive pseudo--interaction, see
(\ref{Pressure1}). This kind of connection between thermodynamics
and the breaking of Lorentz symmetry has not been carried out
\cite{[18], [19], [20]}. This division into two different classes
could be of relevance since the appearance of interactions among
the particles is a necessary, though not sufficient condition, for
the emergence of critical point in thermodynamical systems
\cite{[14]}. In other words, our approach could allow us to
calculate, by means of the cluster expansion, an expression for
the pseudo--interaction, see expression (\ref{2virialcoeff}),
associated to a breaking of Lorentz symmetry, and in this form we
could look for critical points, a fact that has not been
considered previously.

A careful analysis of the link between Lorentz symmetry breaking
and thermodynamics has already shed light upon some of the
characteristics of black--holes \cite{[22]}. This work has proved
that the second law of thermodynamics is strong enough to cope
with the introduction of some of the kind of effects that a
breaking of Lorentz symmetry implies. In this context our work
pursues this analysis for other kind of thermodynamical systems.
The goal behind it is the quest for additional systems that could
be considered amplifiers \cite{[21]}.

Finally, as mentioned at the end of the previous section, a
terrestrial experiment, as a tool to test this kind of violations,
is not a feasible idea. Nevertheless, this last remark does not
imply the uselessness of this approach. Indeed, we may wonder if
this modification could impinge, for instance, in the evolution of
early stages of the universe. As an example of this possibility
let us remember that the theory of nucleosynthesis is inconsistent
with the cluster baryon fraction in a critical density universe
\cite{[23]}. A possible solution to this problem appears if the
theory of nucleosynthesis is flawed, and requires the intervention
of new physics \cite{[24]}. Clearly, a question in this context is
the role that a new thermodynamics could play as part of this new
physics.

\bigskip

\begin{acknowledgments}
 This research was partially supported by CONACYT Grants 48404--F and  47000--F.
 A.C. would like to thank A.A. Cuevas--Sosa for useful
discussions and literature hints.
\end{acknowledgments}

\end{document}